\documentstyle[12pt,epsf]{article}
\begin{document}
\begin{center}
{\Large\bf STUDY OF THE POSSIBILITY OF SUPERNARROW\\
\vspace{0.2cm}
DIBARYON PRODUCTION IN THE $\vec\gamma d\to\pi^{\pm}+D$\\
\vspace{0.2cm}
REACTIONS}
\vspace{0.5cm}

{\large \bf V.V.~Alekseev, S.N.~Cherepnya, L.V.~Fil'kov and
V.L.~Kashevarov}
\vspace{0.5cm}

{\bf Abstract}
\end{center}

The possibility of observation of supernarrow dibaryons, decay
of which into two nucleons forbidden by the Pauli principle,
produced in processes of charged pion photoproduction
by polarized photons from the deuteron is analyzed. It is shown
that the expectable dibaryon yield may exceed the background by 
a factor of 10--100.
\vspace{1cm}

The possibility of existence of multi-quark states is predicted 
by QCD \cite{jaf}. This is a new type of matter. The experimental 
discovery of such states might have serious consequences for both 
elementary particle physics and nuclear physics. A search for 
narrow 6-quark states (dibaryons) is being carried out for a long 
time (see, e.g., \cite{tat,kom}), and a number of candidates
for these states has been found. However, up to the present
day, one can hardly unambiguously state that the features found
in these experiments were indeed dibaryons. This associated,
first of all, with a relatively low contribution of dibaryons
to the processes under study and the uncertainty of the contribution
of the background processes.  

We propose to seek for supernarrow dibaryons whose decay into
two nucleons is forbidden by the Pauli principle 
\cite{filk1,filk2,akh1,akh2,ger,alek}. Such dibaryons
satisfy the condition
\begin{equation}
(-1)^{T+S}P=+1,
\end{equation}
where $T$ is the isotopic spin, $S$ is the internal spin, and $P$
is the parity of the dibaryon. These dibaryons with mass
$M<2m_N+\mu$ ($m_N$ is the nucleon mass and $\mu$ is the pion mass)
can mainly decay into two nucleons with photon emission. The 
contributions of such dibaryons to processes of strong interactions 
of hadrons are very low. However, their contributions to processes
of electromagnetic interactions on light nuclei may be several orders
of magnitude higher than the values of cross sections of these processes 
out of the resonance \cite{filk1,filk2,akh2}.

In the present paper we study the possibility of observation of
supernarrow dibaryons $D(T=1,J^P=1^+,S=1)$ and $D(T=1,J^P=1^-,0)$
having the mass $M<2m_N+\mu$ and satisfying the condition (1) in
the processes of $\pi^{\pm}$ meson photoproduction by polarized
photons from the deuteron 
\begin{equation}
\vec\gamma+d\to\pi^{\pm}+D, D\to\gamma NN .
\end{equation}

The dibaryons under consideration have very small decay widths. Their
values, calculated under assumption that $\gamma NN$ decay of the 
dibaryons goes on mainly through the singlet virtual $^{31}S_0$
level in the intermediate state, are presented in Table 1 
\cite{akh2}.
\begin{table}
\centering
\caption{ Decay widths of $D(1,1^+,1)$ and $D(1,1^-,0)$ dibaryons 
for different dibaryon masses $M$. $\Gamma_t\simeq \Gamma_{\gamma NN}$.}
\begin{tabular}{||c|c|c|c|c|c|c|c||} \hline
$M\;(GeV)$       &1.90&1.91&1.93&1.95&1.98&2.00&2.013 \\ \hline
$\Gamma_t(1,1^+)$&0.2 &0.52&2.2 &5.8 &16  &26  &35    \\
$(eV)$           &    &    &    &    &    &    &      \\ \hline
$\Gamma_t(1,1^-)$&0.05&0.13&0.55&1.46&4   &6.5 &8.75  \\ 
$(eV)$           &    &    &    &    &    &    &      \\  \hline
\end{tabular}
\end{table}  

For the processes (2) the dibaryons can only be produced under condition
when the overlap of the nucleons inside the deuteron is so strong that a 
6-quark state with deuteron quantum numbers is formed. In this case, the 
interaction of a photon or a pion with this state can so change its quantum
numbers that a metastable state satisfying the condition (1) is formed.
Because of this the probability of such dibaryon productions proportional
to the probability $\eta$ of existence of the 6-quark state in the
deuteron. The evaluation of $\eta$ from the difference between the
theoretical and the experimental values of the deuteron magnetic moment
gives $\eta\le 0.03$ \cite{kon}. We assume here that $\eta=0.01$.

Let us consider the photoproduction of dibaryons by photons polarized
at an angle $\alpha$ where
$$
\cos\alpha=\frac{(\vec\epsilon[\vec k_1\ \vec q_1])}{\nu q},
$$
$\vec\epsilon$ and $\vec k_1$ are the plarization and
momentum vectors of an incident photon, $\nu$ is its energy,
$\vec q_1$ and $q$ are the momentum vector of a pion formed in the 
reaction and its absolute value (in lab. system).

Let us restrict our consideration to the dibaryon formation in the
process of $\pi^+$ meson photoproduction. The calculation of
photoproduction of the dibaryons in the combination with $\pi^-$
mesons gives a qualitatively similar result.
The calculations made in the framework of the model \cite{alek}
give the following expression for the $D(1,1^-,0)$ dibaryon production 
cross section in the $\vec \gamma+d\to\pi^++D(1,1^-,0)$ process
(in lab. system):
\begin{eqnarray}
\lefteqn{\frac{d\sigma_{\vec\gamma\to\pi^+D(1,1^-,0)}}{d\Omega}=
\frac23\left(\frac{e^2}{4\pi}\right)\left(\frac{g_1^2}{4\pi}\right)
\eta\frac{q^2}
{m_dM^2\nu J}\left\{|\vec r|^2+\right.} \\ \nonumber
&&\left. q^2\sin^2\theta_{\pi}\sin^2\alpha\left[1-8\frac{m_dr_0}
{\mu^2-t}+4m_d^2\frac{M^2+2r^2_0}{(\mu^2-t)^2}\right]\right\},
\end{eqnarray}
where $t=\mu^2-2\nu(q_0-q\cos\theta_{\pi})$, $J=q(m_d~+\nu)-
q_0\nu\cos\theta_{\pi}$, $m_d$ is the deuteron mass, $q_0(q)$
is the pion energy (momentum),
$$
q_0=\frac1{c_1}\left[(m_d+\nu)c_2\pm\nu\cos\theta_{\pi}
\sqrt{c_2^2-2\mu^2c_1}\right],
$$
$$
c_1=2[(m_d+\nu)^2-\nu^2\cos^2\theta_{\pi}], \; c_2=s+\mu^2-M^2 .
$$
The dibaryon energy $r_0$ and momentum $|\vec r|$ are given by
$r_0=m_d+\nu-q_0$ and $|\vec r|=\sqrt{r_0^2-M^2}$, respectively.

The $D(1,1^+,1)$ dibaryon production cross section has the form
\begin{eqnarray}
\lefteqn{\frac{d\sigma_{\vec\gamma d\to\pi^+D(1,1^+,1)}}{d\Omega}=
\frac{16}{3}\left(\frac{e^2}{4\pi}\right)\left(\frac{g^2_2}{4\pi}
\right)\eta\frac{q^2}{m_dM^2\nu J}\left\{M^2+\right.} \\ \nonumber
&&q^2\sin^2\theta_{\pi}\sin^2\alpha\left.\left[1-4\frac{m_dr_0}
{\mu^2-t}+4\frac{m^2_d|\vec r|^2}{(\mu^2-t)^2}\right]\right\}.
\end{eqnarray}
The values of $g_1^2/4\pi$ and $g_2^2/4\pi$ are unknown. These are 
the constants of strong interaction. Let us set them equal to 1
lest the cross section values should be overstated. To evaluate
the contribution of the dibaryons for different values of the mass
$M$, we assume the possibility of the existence of the dibaryons 
with the masses $M$=1.9, 1.95, and 2.00 $GeV$.

We carried out a numerical analysis of the cross section for 
photoproduction of the dibaryons by photons polarized parallel
($\alpha=90^{\circ}$) and perpendicular $(\alpha=0^{\circ})$
to the reaction plane. As follows from the calculations, the
$D(1,1^-,0)$ dibaryon photoproduction cross section has a strong 
dependence on the photon polarization. For the photons polarized 
in the reaction plane it is large and, at least in the range of 
angles $\theta=10^{\circ}-50^{\circ}$, substantially exceeds the
$D(1,1^+,1)$ dibaryon photoproduction cross section, whereas
the cross section for the photons polarized perpendicular to the 
reaction plane is very small.

The dependence of the $D(1,1^+,1)$ dibaryon photoproduction cross 
section on the photon polarization is considerably weaker. This is
caused by smallness of the factor $|\vec r|^2$ in the latter term
of expression (4).  

The CLAS detector of charged and neutral particles in combination 
with the tagging system of the CEBAF accelerator offer perfect
potentialities for the search and study of supernarrow dibaryons
in the reactions under consideration. This setup enables one to
seek for the dibaryons by way to detecting charged pions with 
a wide-aperture magnetic spectrometer and separating out peaks in 
the spectrum of missing masses. An additional detection of the photon 
produced through the dibaryon decay, in coincidences with the pion,
provides a substantial suppression of the contribution of background
reactions to this spectrum.

The main background reaction in which a photon is produced in the 
final state are the photoproduction of two pions
$(\gamma d\to\pi++\pi^0+nn, \; \pi^0\to\gamma \gamma)$ and the 
radiative photoproduction of a $\pi^+$ meson
$(\gamma d\to\pi^++\gamma+nn)$. When estimating the contribution of
the background reactions, we assumed that they had only a weak 
dependence on the photon polarization.

Using the Monte-Carlo method, we performed simulation of the dibaryon
photoproduction and the main background reactions on the deuteron 
target 20 $cm$ long on the setup mentioned above under real 
conditions. The energy of the initial photons was took in the range
of 500--1000 $MeV$.

Fig.1 illustrates the spectrum of missing masses which was obtained
in this simulation for 5 hours of the accelerator work. The expected
yields of the dibaryons with different masses are listed in Table 2.

\begin{table}
\centering
\caption{ Expected yields of the dibaryons produced in the
$\vec\gamma d\to\pi^+D$ process during 5 hours of accelerator work.}
\begin{tabular}{||c|c|c|c|c||} \hline
         &$\alpha$&$M=1.90\,GeV$&$M+1.95\,GeV$&$M=2.00\,GeV$ \\ \hline
$D(1,1^+,1)$& $90^{\circ}$& 386 &   370       &  351    \\
            & $0^{\circ}$ & 443 &   437       &  381    \\ \hline
$D(1,1^-,0)$& $90^{\circ}$& 3901&   3552      &  3165   \\
            & $0^{\circ}$ &   7 &     5       &    3    \\ \hline
\end{tabular}
\end{table}

Among two background processes considered, the contribution to the 
mass range $M<2m_N+\mu$ being studied is give by the radiative 
photoproduction only.

As seen from Fig.1 and Table 2 the contribution of the dibaryons 
may exceed the background by a factor of 10--100. 
For $\alpha=90^{\circ}$, the yield of $D(1,1^-,0)$ dibaryons must
exceed the yield of $D(1,1^+,1)$ dibaryons by about a factor of ten.
For $\alpha=0^{\circ}$, the production of $D(1,1^-,0)$ dibaryons is
strongly suppressed (a 5-hour exposure is expected to give only a few
events). From the comparison with the dibaryon production
 by unpolarized photons it follows that the excess of the yield
of $D(1,1^-.0)$ dibaryons produced by polarized photons over the
background is several times greater than the corresponding excess for 
unpolarized photons.

Thus, the use  of a polarized photon beam makes it possible to increase
the contributions of the dibaryons to the mass spectrum in comparison
with the contribution of the background and to determine the dibaryon 
quantum numbers. Additional information on these quantum numbers can
be obtained from the analysis of the angular distributions of the 
differential cross sections for the dibaryon photoproduction.

If the dibaryons are detected in the process of $\pi^+$ meson 
photoproduction from the deuteron, they will have to be observed in the
$\vec\gamma d\to\pi^-D$ reaction as well. The observation of dibaryons
in both reactions will allow to make an conclusion about the production
of the supernarrow dibaryons satisfying the condition (1) more
unambiguously. Moreover, this will enable one to determine the possible
electromagnetic splitting of their masses and obtain additional
conditions for determining the dibaryon quantum numbers.

This work was supported by the Russian Foundation for Basic Research,
project No. 96-02-16530A.

\normalsize
\pagestyle{plain}
\setcounter{figure}{0}
\newpage
\begin{figure}[htp]
\hspace*{-1cm}
\vspace*{-2cm}
\epsfxsize=18.5cm
\epsfysize=20cm
\epsffile{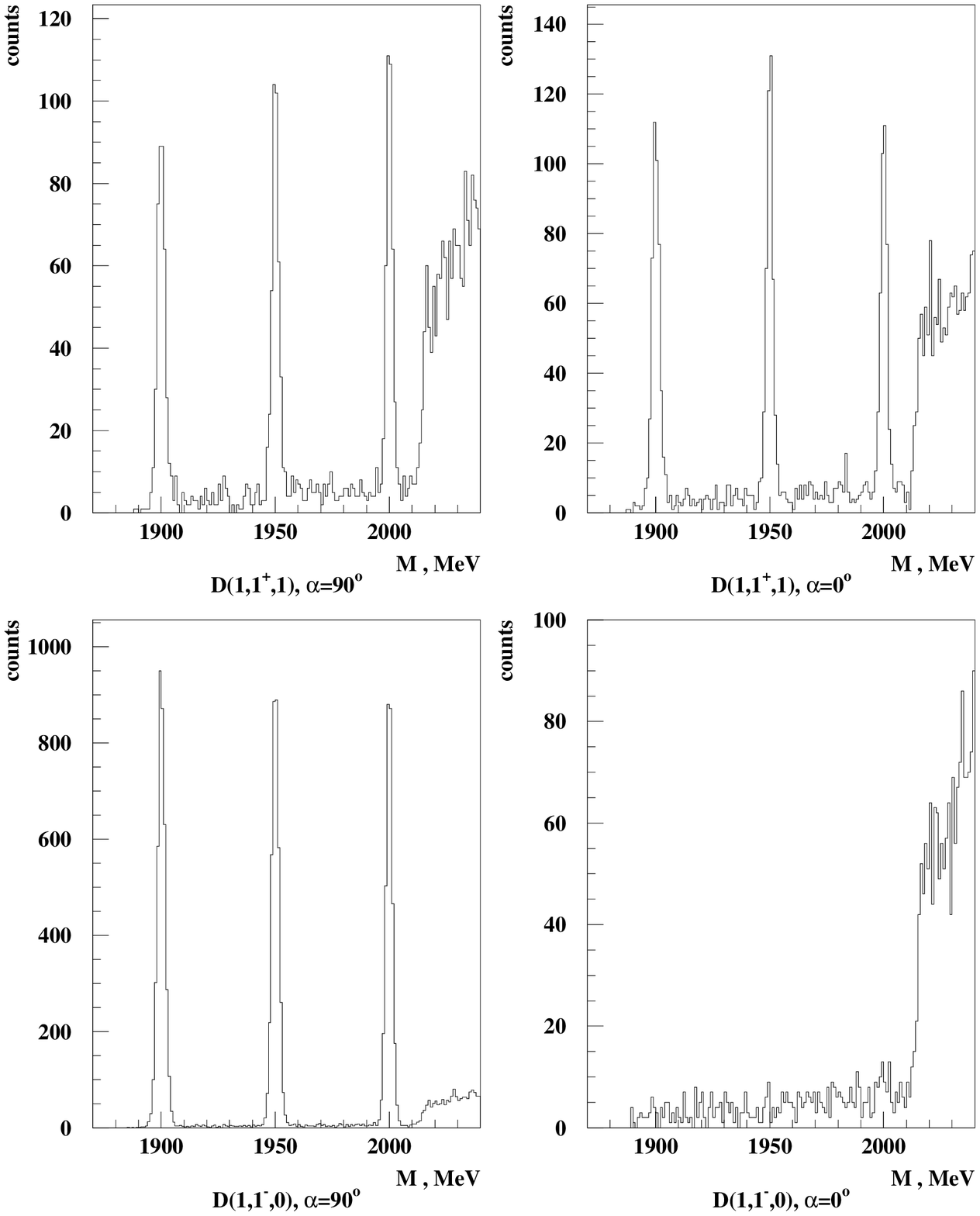}
\caption{ Spectra of missing masses expected for the $D(1,1^+,1)$ and
$D(1,1^-,0)$ dibaryons produced by photons polarized at the angles
$\alpha=0^{\circ}$ and $90^{\circ}$ from the deuteron and for the
main background processes.}
\end{figure}

\end{document}